# Evolutionary search for superhard materials: Methodology and applications to forms of carbon and TiO2

Andriy O. Lyakhov*, Artem R. Oganov*,†

* *Department of Geosciences, and Department of Physics and Astronomy, Stony Brook University, Stony Brook 11794-2100, NY, USA*
† *Geology Department, Moscow State University, 119992, Moscow, Russia*

## Abstract

We have developed a method for prediction of the hardest crystal structures in a given chemical system. It is based on the evolutionary algorithm USPEX (Universal Structure Prediction: Evolutionary Xtallography) and electronegativity-based hardness model that we have augmented with bond-valence model and graph theory. These extensions enable correct description of the hardness of layered, molecular, and low-symmetry crystal structures. Applying this method to C and TiO2, we have (i) obtained a number of low-energy carbon structures with hardness slightly lower than diamond and (ii) proved that TiO2 in any of its possible polymorphs cannot be the hardest oxide, its hardness being below 17 GPa.

Hardness is an important property that determines many of the technological applications of materials. The ability to predict the hardest phase of a crystal, given its stoichiometry, will enable a systematic search for novel hard materials and allow one to appraise controversial experimental results. The problem at hand is in principle, similar to the search for the most stable phase at given conditions – an area where many interesting results were achieved recently (for an overview, see [1]). Both problems require finding a global minimum on some surface in multidimensional space: hardness and free energy, respectively, as a function of the atomic positions and lattice vectors. However, computing the hardness of a given crystal structure is a difficult task. Only recently was significant progress achieved in formulating models that are physically meaningful and sufficiently accurate for several classes of compounds, see [2-5]. Building up on these developments, we present an evolutionary algorithm that searches for the hardest possible structure of a given material. It involves the concept of *hybrid global optimization*, where global optimization with respect to the hardness is conducted in the space of local minima of the (free) energy. This algorithm opens up a new way of materials design and discovery, as exactly the same approach can be used for optimizing other properties.

Our approach is based on the evolutionary algorithm USPEX [6-8], but in this study the fitness function was chosen to be hardness instead of the free energy. In this approach, a set of candidate solutions (*population*) evolves, driven by *selection* and modified by certain rules (*variation operators*) [7]. In our case, a candidate solution is a locally optimized structure described by atomic positions and unit cell vectors. Local optimization (i.e. structure relaxation) should be done with respect to the energy, rather than hardness, to ensure that the structure is chemically realistic. *Selection* process eliminates 'bad' candidates (the 'quality' is determined by the value of the fitness function).

For an overview of analytical models of hardness based on crystal structure, see Refs. [5] and [9]. Here we build up on the model of Li *et al* [4], which computes the hardness based on the electronegativities and covalent radii of the constituent atoms, and the bond lengths in the structure. Like most of the other models [2, 3], it gives good results for simple high-symmetry structures. For low-symmetry and/or anisotropic structures, and in particular for molecular and layered structures, the results are much less satisfactory. This poses a serious problem for global optimization, because the majority of structures produced during the evolutionary search are rather complex and have low symmetry. For such structures, also the concept of well-defined integer coordination number, as used in [4], is often inadequate. As we show, these deficiencies can be remedied.

We have generalized the approach of Li *et al* so as to correct the above mentioned pathologies while reproducing the results of the original model [4] for 'good' cases. Let us denote by $n$ the number of different bond types in the unit cell, and we label these with the index $k=[1,...,n]$. The model [4] computes electronegativities of each $i$-th atom as $\chi = 0.481 n_i/R_i$, where $n_i$ and $R_i$ are the number of valence electrons and univalent covalent radius of this atom, respectively. To take into account the dependence of the electronegativity on the environment, and deviations of actual bond lengths $R_k$ from the sum of covalent radii, we correct the electronegativites of atoms $i$ and $j$ participating in the bond $k$

$$\chi_i^k = 0.481 \frac{n_i}{R_i + \Delta_k/2} \qquad \chi_j^k = 0.481 \frac{n_j}{R_j + \Delta_k/2} \qquad (1)$$

by equally distributing $\Delta_k = R_k - R_i - R_j$ (in Å) between the bonded atoms. This introduces explicit dependence of the electronegativity and hardness on bond lengths. The 'effective coordination number' that describes the atomic valence involved in each bond is defined as $CN_i^k = v_i/s_i^k$, where $v_i$ is the valence of atom $i$ (in general not equal to $n_i$) and $s_i^k$ is a bond valence that can be calculated using Brown's model [10]:

$$s_i^k = \frac{v_i \exp(-\Delta_k/0.37)}{\sum_{k'} \exp(-\Delta_{k'}/0.37)},$$

here the sum goes over all bonds $k'$ in which atom $i$ participates. This definition involves renormalization to satisfy exactly the sum rule [10] $\sum_{k'} s_i^{k'} = v_i$. Note that $CN_i^k$ is a continuous function of structure and can take non-integer values (unlike classical coordination numbers), which is very useful for global optimization. Now we substitute these generalized formulas into the original formulas [4] for the hardness. Average electron-holding energy of the bond and its ionicity indicator are defined as in the original work of Li *et al.* [4]

$$X_k = \sqrt{\frac{\chi_i^k \chi_j^k}{CN_i^k CN_j^k}} \qquad f_k = \frac{|\chi_i^k - \chi_j^k|}{4\sqrt{\chi_i^k \chi_j^k}}, \qquad (2)$$

which are then substituted into the formula [4] for the Knoop hardness (in GPa):

$$H = \frac{423.8}{V} n \left[ \prod_{k=1}^{n} N_k X_k e^{-2.7 f_k} \right]^{1/n} - 3.4, \qquad (3)$$

where $V$ is the volume of the unit cell and $N_k$ is the number of bonds of the type $k$ in the unit cell. Coefficients 423.8, 2.7 and -3.4 were obtained in [4] by fitting to experimental data for hard materials.

From eq. (3) one can see that the hardness (as a geometric average) is strongly affected by the weakest included bonds. The case of graphite is very instructive. Using the standard coordination number (three) includes only the strong covalent bonds in eq. (3). Their lengths are 1.42 Å and in the original model their electron-holding energy is equal to $X = 0.844$. With the unit cell volume $V = 35.28$ Å$^3$ one obtains an unrealistically high $H = 57$ GPa. The original model [4] takes into account the number and strength of the bonds in the structure, but not the structural topology. Yet, the real hardness of graphite is determined by the weak van der Waal bonds between the layers of the structure. Thus, by *hardness-defining* (or structure-forming) bonds we mean not only the strongest chemical bonds, but also a set of bonds necessary to maintain thee-dimensionality of the crystal structure. These bonds need to be included in eq. (3) and we discovered an automatic way to find them. Let us describe the crystal as a graph where atoms are vertices and *hardness-defining* bonds are edges. The challenge for the algorithm is to determine edges knowing only the geometric arrangement of atoms and their chemical identities. We do this by gradually adding to the graph those weak bond groups that decrease the number of its connected components. In the case of graphite, strong sp$^2$ bonds within the layer and the closest bonds between the layers are determined as *hardness-defined* ones. Weak bonds have the length of 3.35 Å, their electron-holding energy is $X = 0.002$. Geometric average in (3) results in a hardness $H = 0.17$ GPa, in agreement with our everyday experience that graphite is an ultrasoft solid and a lubricant.

The complete connectivity of the graph is a sufficient but not necessary condition for determining whether all *hardness-defining* bonds are taken into account. There is one important general case where a disconnected graph will still represent a 3D-bonded structure. A simple illustration of this phenomenon is a 3D chess board, where all white and black cubes build their own connected subgraphs and these subgraphs are not connected with each other. Amazingly, such exotic structure is known in nature: it is the structure of the "3D catenane", cuprite ($Cu_2O$). Such non-connected but intersecting graphs can be detected using multi-color graph theory, which we have implemented.

With these extensions (bond valence model and graph theory) the model of Li et al. [4] shows excellent performance without the need for changing the final formulas or refitting the coefficients, as illustrated in **Table I**. It is worth noting that our values of the hardness correspond to experimental microhardness (hardness in the limit of non-plastic deformations) measured by Knoop hardness test. It may be surprising that a model based on the ideal crystal structure so closely reproduces the experimental microhardness, which significantly depends on defects (especially dislocations) [11]. Our understanding is that softening due to dislocations is factored in the fitted coefficients in eq. (3). This model takes into account the most important chemical effects related to strength of covalent bonding, degree of ionicity and directionality, and topology of the crystal structure. It seems to be applicable even to very complex cases. For example, hardnesses obtained for different phases of boron are 41.0 GPa for α-$B_{12}$, 38.0 GPa for β-$B_{106}$ and 42.5 GPa for γ-$B_{28}$, which compare quite well with the experimental Vickers hardness values of 42 GPa, 45GPa and 50 GPa respectively [12 and references therein].

**TABLE I.** Hardness of different materials (in GPa).

| Material[13] | Model [4] | Present work | Experiment[14] |
|---|---|---|---|
| Diamond | 91.2 | 89.7 | 90[15] |
| Graphite | 57.4 | 0.17 | 0.14[16] |
| Rutile, $TiO_2$ | 12.9 | 14.0 | 8-11[17] |
| $TiO_2$ Cotunnite | 16.6 | 15.3 | controversial |
| $\beta$-$Si_3N_4$ | 23.4 | 23.4 | 21[18] |
| Stishovite, $SiO_2$ | 31.8 | 33.8 | 32[19] |

This model of hardness, coupled with a global optimization algorithm for crystal structure prediction, can be used for predicting the hardest structure in a given chemical system. We applied the method to two particularly interesting systems, carbon and $TiO_2$. These cases address major problems in the field of superhard materials. In particular, does a material harder than diamond exist? Other carbon allotropes (e.g., see [20]) are prime suspects here. The search for the hardest oxide is another important problem: while diamond burns in the oxygen atmosphere at high temperatures, oxides can be inert to oxygen, which could make them important for manufacturing novel types of cutting and abrasive tools. Previous proposals included stishovite [19], $TiO_2$-cotunnite [21] and $B_6O$ [22]. While boron suboxide $B_6O$ is the hardest known oxide ($H$ = 45 GPa [22]), its thermal stability in the oxygen atmosphere is rather poor [23]. Other ultrahard oxides are stishovite with $H$ = 32 GPa [19], seifertite (high-pressure polymorph of $SiO_2$, predicted [5] to be slightly harder than stishovite) and $TiO_2$-cotunnite with the reported, but controversial, hardness $H$ = 38 GPa [21]. Here we investigate whether $TiO_2$, in any of its forms, can be as hard as reported.

In our structure searches, hardness was evaluated for structures after they were relaxed. Relaxation was done using the generalized gradient approximation [24] as implemented in the VASP code [25] for C and $TiO_2$, and in additional searches using the GULP code [26] and a Buckingham potential [27] for $TiO_2$. All ab initio calculations used plane-wave basis sets and Monkhorst-Pack meshes for Brillouin zone, sufficient for excellent convergence in the total energy and stress tensor. For each system we explored different numbers of formula units in the unit cell (up to 24 atoms/cell). A typical result is shown in **Fig. 1**.

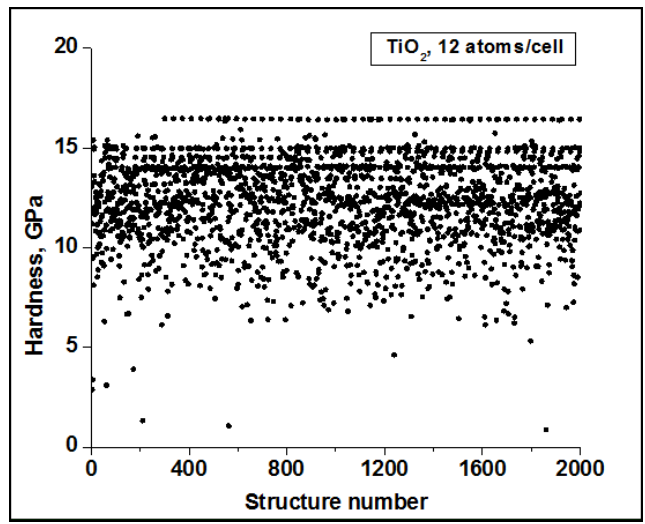

**Fig. 1.** Example of evolutionary hardness optimization: an *ab initio* run for $TiO_2$.

For carbon, the hardest structure found is diamond. Lonsdaleite and polytypes intermediate between diamond and lonsdaleite also emerged in our runs, but they are slightly softer. Our searches at various system sizes (in total, we sampled 9500 structures) produced a large number of superhard allotropes with hardnesses approaching that of diamond, see **Table II** and **Fig. 2**. Parameters of these structures can be found in supplementary material [28]. Synthesis and practical applications of some of these structures may be possible. Indeed, there are indications that the *C2/m* or *I/4mmm* structure has been obtained on cold compression of graphite [29, 30] (recently, the C2/m structure got additional theoretical support [32]). We also note that while diamond is the hardest possible phase of carbon, it is not the densest one. Several significantly denser (and only marginally softer) metastable structures were proposed in Ref. [33].

**TABLE II.** Some of the superhard allotropes of carbon found by USPEX. Novel structures are described by symmetry group and number of atoms in the unit cell. Hardness is measured in GPa, volume - in Å$^3$/atom, enthalpy - in eV/atom relative diamond.

| Structure | H | Enthalpy | V |
|---|---|---|---|
| Diamond | 89.7 | 0.000 | 5.685 |
| Lonsdaleite | 89.1 | 0.026 | 5.696 |
| M-carbon (*C2/m*) [6, 29] | 84.3 | 0.163 | 5.969 |
| bct4-carbon (*I4/mmm*) [30,31] | 84.0 | 0.198 | 6.003 |
| *Cmcm* – 16 | 83.5 | 0.282 | 6.036 |
| *P2/m* – 8 | 83.4 | 0.166 | 5.979 |
| *I2$_1$2$_1$2$_1$* – 12 | 82.9 | 0.784 | 5.911 |
| *Fmmm* – 24 | 82.2 | 0.322 | 6.128 |
| *Cmcm* – 12 | 82.0 | 0.224 | 6.157 |
| *P6$_5$22* [34] | 81.3 | 0.111 | 6.203 |

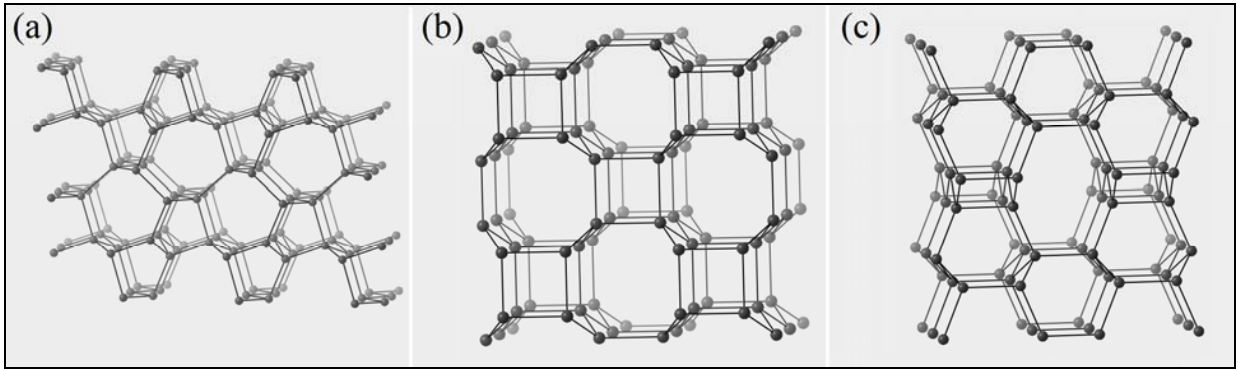

**Fig. 2.** Predicted superhard carbon allotropes. (a) M-carbon, (b) bct4-carbon and (c) *Cmcm*-16 structure.

Our numerously repeated simulations of $TiO_2$ consistently indicated that the reported hardness of 38 GPa [21] is extremely unlikely to be correct. The highest possible hardness for a $TiO_2$ polymorph is 16.5 GPa (from *ab initio* calculations) or 15 GPa (classical force field). The "ultrahard" $TiO_2$-cotunnite with $H = 38$ GPa is thus an artifact; our model gives $H = 15.3$ GPa for it, see **Table I**. In this structure each Ti atom has nine bonds with O atoms, their lengths ranging from 2.03 Å to 2.56 Å (with electron-holding energy from 0.43 to 0.02 respectively). Ionicity indicator from (2) is in the range 0.31-0.38. One can see from (2) and (3) that the relatively low hardness of TiO$_2$-cotunnite is caused by the high coordination number and relatively high ionicity. Also, theoretical calculations [35] suggest that this structure is dynamically unstable at 1 atm and careful measurements of the equation of state [36, 37] showed that Dubrovinsky's measurements [21] overestimated the bulk modulus by about 40%. Therefore the experimental data [21] need to be reconsidered. The hardest chemically feasible TiO$_2$ structures with 12 atoms in the unit cell obtained by us can be divided into three groups: (i) structures related to rutile ($H \sim$ 14 GPa); (ii) structures related to *Pca2$_1$* orthorhombic structure (7-9% denser than rutile with energies about 0.03-0.08 eV/atom higher); (iii) structures related to the *Pnma* orthorhombic structure with $H$ = 15.7 GPa, depicted in **Fig. 3**. Details of some low-energy hard TiO$_2$ structures can be found in supplementary material [28].

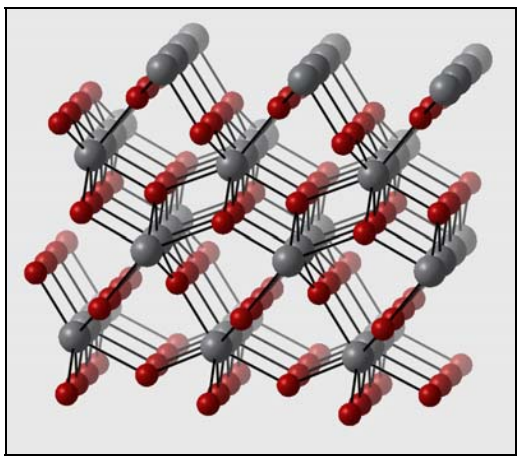

**Fig. 3.** Hardest low-energy structure of TiO$_2$ found by USPEX. Its enthalpy is 0.1 eV/atom above the ground state, while the density is 11% higher.

We have also done preliminary calculations for C$_3$N$_4$, a material some structures of which were predicted to be less compressible [38, 39] and potentially harder than diamond. None of the structures that we have sampled are harder than diamond, but those structures proposed in Ref. [40] come close to diamond in hardness. Cubic C$_3$N$_4$ has theoretical hardness $H$ = 86.9 GPa, pseudocubic C$_3$N$_4$ has $H$ = 84.6 GPa, β-C$_3$N$_4$ has $H$ = 79.9 GPa – these numbers are consistent with [4]. Our global searches did not indicate any structures of C$_3$N$_4$ with hardnesses above 87 GPa. Thorough variable-composition runs in the C-N system, looking for the hardest material in this system, have also arrived at pure carbon in the diamond structure.

In summary, we proposed an improved empirical model to predict the hardness of materials based only on their crystal structure. Merged with evolutionary crystal structure prediction algorithm USPEX, it provides a way of systematic discovery of new hard materials. Our results show that diamond is the hardest carbon allotrope and that all possible TiO$_2$ polymorphs are relatively soft (H < 17 GPa). Thus TiO$_2$ cannot be considered as one of the hardest oxides, resolving a long-standing controversy. We have found several sp$^3$-allotropes of carbon with very high hardnesses, simple structures and reasonably low energies. These may be synthesizable. The same concept of hybrid global optimization as used here for hardness, can be used for optimizing other properties of materials, and in near future will become the basis of computational materials discovery and design.

We gratefully acknowledge funding from Intel Corp., DARPA (grant N66001-10-1-4037), National Science Foundation (EAR-1114313) and Rosnauka (Russia, contract 02.740.11.5102). Calculations were performed on the Blue Gene supercomputer at New York Center for Computational Sciences, on the Skif MSU supercomputer (Moscow State University), and at Joint Supercomputer Centre of the Russian Academy of Sciences. USPEX code, with options for global optimization of the thermodynamic potential (energy, enthalpy, free energy), density, hardness, and other properties, is available at: http://han.ess.sunysb.edu/~USPEX.